\documentclass[12pt]{article}
\usepackage{authblk}
\usepackage{here}
\usepackage{xcolor}
\usepackage{ulem}
\usepackage{amsfonts}
\usepackage{amsmath,amssymb}
\usepackage{mathrsfs}
\usepackage{bm}        
\usepackage{graphicx}          
\usepackage{ascmac}
\usepackage{braket}
\usepackage{cite}
\usepackage[top=25truemm,bottom=30truemm,left=20truemm,right=20truemm]{geometry}
\usepackage{afterpage}
\usepackage{cases}
\usepackage{url}
\usepackage{multirow}
\numberwithin{equation}{section}
\usepackage[linktocpage=true]{hyperref}

\hypersetup{
  linktoc = all,
  colorlinks   = true,
  linkcolor = blue,
  anchorcolor = black,
  citecolor = blue,
  filecolor = cyan,
  menucolor = red,
  runcolor = cyan,
  urlcolor = magenta,
}

\newcommand{\cF}{{\mathcal F}}

\newcommand{\cL}{{\mathcal L}}
\newcommand{\cM}{{\mathcal M}}
\newcommand{\cN}{{\mathcal N}}

\newcommand{\bZ}{{\mathbb Z}}

\newcommand{\ha}{{\hat{a}}}
\newcommand{\hb}{{\hat{b}}}
\newcommand{\hd}{{\hat{d}}}

\newcommand{\hu}{{\hat{u}}}

\newcommand{\hst}{{\hat{\star}}}

\newcommand{\nn}{\nonumber}

\newcommand{\ol}{\overline}

\newcommand{\wt}{\widetilde}

\renewcommand{\em}{\it}

\renewcommand{\cal}[1]{\mathcal{#1}}

\renewcommand{\rm}[1]{\mathrm{#1}}

\newcommand{\lrC}[1]{{\left(#1\right)}}

\newcommand{\M}{{\cal{M}_6}}

\newcommand{\GA}{{A_1}}

\newcommand{\GB}{{B_2}}

\newcommand{\Gb}{{B^{\rm{CW}}_2}}
\newcommand{\GC}{{C^{\rm{CW}}_3}}
\newcommand{\GD}{{D^{\rm{CW}}_4}}
\newcommand{\GE}{{E^{\rm{CW}}_5}}
\newcommand{\Sb}{{G_3}}
\newcommand{\SC}{{H_4}}
\newcommand{\SD}{{I_5}}
\newcommand{\SE}{{J_6}}

\newcommand{\Jb}{{j_{\rm{CW}}^{[1]}}}
\newcommand{\JC}{{j_{\rm{CW}}^{[2]}}}
\newcommand{\JD}{{j_{\rm{CW}}^{[3]}}}
\newcommand{\JE}{{j_{\rm{CW}}^{[4]}}}

\title{{\LARGE Note on higher-group structure in 6d self-dual gauge theory}}

\author[1]{Tatsuki Nakajima}
\author[1]{Kikyo Nakamura\thanks{present address: Meitetsucom Co., Ltd.}}
\author[2,1]{Tadakatsu Sakai}
\affil[1]{Department of Physics, Nagoya University}
\affil[2]{Kobayashi-Maskawa Institute for the Origin of Particles and the Universe, Nagoya University}

\date{}
\begin{document}
\maketitle

\begin{abstract}
We analyze higher-group structure of a 6d model coupled with a self-dual
2-form gauge field. This model is defined from 6d 
axion-electrodynamics with a 1-form Chern-Weil(CW) symmetry gauged
dynamically. The gauging leads to a Green-Schwarz-West-Sagnotti(GSWS) 
term, which gives rise to an anomaly through
a GSWS transformation acting on the 2-form gauge field.
We cancel this anomaly by gauging a 3-form CW symmetry
in 6d axion-electrodynamics.
We find out the global symmetries in the resultant model 
and derive the gauge invariant action in the presence
of the background gauge fields. It is argued that a discrete 1-form
symmetry is anomalous because turning on the associated
background gauge field causes quantum inconsistency
due to an operator-valued ambiguity. 
Higher-group structure in this model that is manifested as a
Green-Schwarz-like transformation 
for CW background gauge fields is discussed.

\end{abstract}
  
\tableofcontents
\section{Introduction}

The paper \cite{Nakajima:2022feg} discusses higher-group structure
in 6d axion-electrodynamics by extending interesting papers
\cite{Hidaka:2020iaz,Hidaka:2020izy}, which explore
3-group structure encoded in 4d axion-electrodynamics.
It is seen that the 6d model possesses much richer structure of
higher-group compared with the 4d case, because the 6d model
allows Chern-Weil(CW) currents of higher rank to exist.
The higher-group structure is manifested as a Green-Schwarz(GS)-like
transformation law \cite{Green:1984fu}
for the background gauge fields associated
with the CW symmetry, although the mathematical framework for it is 
difficult to formulate.
{}For recent studies of higher-group structure, see also 
\cite{Kapustin:2013uxa,Kapustin:2013qsa,Sharpe:2015mja, Bhardwaj:2016clt,Kapustin:2017jrc,Tachikawa:2017gyf, deAlmeida:2017dhy, Benini:2018reh,Cordova:2018cvg, Delcamp:2018wlb, Wen:2018zux, Delcamp:2019fdp, Thorngren:2020aph,Cordova:2020tij,Hsin:2019fhf,Hsin:2020nts,Gukov:2020btk,Iqbal:2020lrt,Brauner:2020rtz,DeWolfe:2020uzb,Heidenreich:2020pkc,Apruzzi:2021vcu,Bhardwaj:2021wif,Hidaka:2021mml,Hidaka:2021kkf,Apruzzi:2021mlh,DelZotto:2022joo,Cvetic:2022imb,Nakajima:2022jxg,Kan:2023yhz,Nakayama:2023rtn}.

In this paper, we discuss higher-group structure in a 6d model with
a self-dual gauge field.
This model is obtained by promoting a CW 2-form background gauge field
in the 6d axion-electrodynamics to a dynamical gauge field.
One of the motivations for studying it is 
that this serves as a toy model that is simple enough to analyze
and encodes an interesting higher-group
structure as in the original 6d axion-electrodynamics.
{}Furthermore, by examining the algebraic structure of the symmetry 
generators using the same procedures as in
\cite{Hidaka:2020iaz,Hidaka:2020izy,Nakajima:2022feg},
it is expected that we make a key step toward a deeper understanding
of the mathematical foundations of higher-group. In this paper, 
we leave it for a future work to explore the algebraic structure in detail.

The gauging of the CW symmetry leads to a coupling
between the 2-form self-dual gauge field $b$ and the CW 4-form current
$\frac{1}{4\pi^2}da\wedge da$ with $a$ being a photon field.
This is equal to a Green-Schwarz-West-Sagnotti(GSWP) term
\cite{Green:1984bx,Sagnotti:1992qw}, which enables us
to cancel 6d reducible anomalies by using a GS-like
transformation law obeyed by $b$.
A typical example is given by a $U(1)$ gauge theory coupled with
chiral fermions, 
where the box $U(1)^4$ anomaly can be canceled  
by using the GSWP mechanism.
In this paper, we focus on an anomaly free model
that is constructed by gauging another CW symmetry in the 6d 
axion-electrodynamics with no chiral fermions coupled. 
As discussed in \cite{Nakajima:2022feg},
the 6d axion-electrodynamics admits the CW current $da/2\pi$.
We gauge this symmetry by introducing a dynamical 4-form gauge field
coupled with the CW current.
The anomaly free model is obtained by requiring that the 4-form
gauge field make an appropriate GS-like transformation
under the $U(1)$ gauge transformation as well.

We next discuss the global symmetries realized in the
resultant 6d model. It is seen that a discrete 1-form symmetry found
in \cite{Nakajima:2022feg} that acts on the photon is broken explicitly,
but gets recovered by modifying it so that it acts on
the self-dual gauge field and 4-form gauge field 
together with the photon.
We argue that the 1-form symmetry is anomalous, however, because turning on
the corresponding background field results in an operator-valued ambiguity,
which states that the theory in the presence of the background
gauge field has a quantum Chern-Simons(CS) action that is 
ambiguous in lifting it to 7d.
It is found that higher-group structure in this model is less
interesting than that in 6d axion-electrodynamics generically,
because the GS-like transformation is induced only for a 1-form
CW background gauge field for generic cases.

The organization of this paper is as follows. In section 2, we construct
the 6d model by gauging dynamically CW symmetries in 6d axion-electrodynamics.
Section 3 works out the global symmetries of the model by studying
the equations of motion in detail.
In section 4, we turn on the background gauge fields for the global
symmetry to show that an operator-values ambiguity is caused by 
that for the discrete 1-form symmetry.
Higher-group structure in this model is discussed in section 5.
The appendix A discusses dimensional reduction of the 6d model
to 5d for the purpose of evading the subtlety of the self-dual
2-form gauge field in obtaining the global higher-form symmetries
in 6d.

\vspace{.2cm}
\noindent
{\bf Notation}

In this paper, dynamical fields to be path integrated are 
denoted with lowercase letters. We also have 
background gauge fields that are necessary to promote
a global symmetry to a local symmetry.
These fields are denoted with capital letters.

\section{Model}
The action of the 6d axion electrodynamics is given by
\begin{align}
  S_{\rm{aE}}[\phi,a]=-\int_{\M}\lrC{\frac{1}{2}|d\phi|^2+\frac{1}{2}|da|^2-\frac{N}{48\pi^3}\phi(da)^3} \ .
\end{align}
Here, $|\chi_p|^2=(1/p!)\,\chi_p\wedge\star\chi_p$ for the 
$p$-form $\chi_p$ with
$\star$ being the Hodge star.
We take $\cM_6$ to be a spin manifold so that $N\in\bZ$. We denote
$(da)^3=da\wedge da\wedge da$ for simplicity.
This action admits discrete $\bZ_N$ symmetries that follow from
the equation of motion(EoM) for $\phi$ and $a$.
$a$ is normalized as $\int da\in2\pi\bZ$.
Turning on the corresponding background field gives rise to
an operator-valued ambiguity. This is canceled by
gauging the CW symmetry and requiring that
the associated background gauge field make an appropriate
GS transformation. The CW current reads\footnote{
{}For the normalization of the CW currents, we follow that of 
\cite{Cordova:2020tij}, where the integral of the CW current over a closed 
cycle in $\cM_6$ is $\bZ$-valued.}
\begin{align}
  \Jb=\frac{1}{4\pi^2}\,(da)^2\ ,\qquad
  \JC=\frac{1}{4\pi^2}\,d\phi\wedge da\ ,\qquad&
  \JD=\frac{1}{2\pi}\,da\ ,\qquad
  \JE=\frac{1}{2\pi}\,d\phi \ .\nn
\end{align}
Let $B_2^{\mathrm{CW}},C_3^{\mathrm{CW}},D_4^{\mathrm{CW}}$ and 
$E_5^{\mathrm{CW}}$
be the background gauge fields that couple minimally with
the CW currents $\Jb,\JC,\JD$ and $\JE$ respectively.
It is shown that the gauge invariant field strength for
the CW gauge field should be defined by
\begin{align}
\begin{split}
  \Sb&=d\Gb-\frac{N}{2\pi}\GA\GB\ ,\\
  \SC&=d\GC+\frac{N}{4\pi}(\GB)^2\ ,\\
  \SD&=d\GD-\frac{1}{2\pi}\GA d\GC+\frac{1}{2\pi}\GB d\Gb-\frac{N}{4\pi^2}\GA(\GB)^2\ ,\\
  \SE&=d\GE+\frac{N}{2\pi}\GB d\GC+\frac{N}{12\pi^2}(\GB)^3\ .
\end{split}  
\label{FS}
\end{align}
Here, $(A_1,A_0)$ with $NA_1=dA_0$ and $(B_2,B_1)$ with $NB_2=dB_1$
are the background $\bZ_N$ gauge fields for the EoM-based 0-form and
1-form $\bZ_N$ symmetries, respectively.
Hereafter, we denote a symmetry group $G$ of rank $p$ as
$G^{[p]}$. 
{}For instance, the $U(1)$ gauge symmetry acting on the photon
field is a 0-form symmetry and is referred to as $U(1)^{[0]}$
with the gauge transformation given by $\delta^{[0]}a=d\lambda_0$.

Now we promote $C_{2}^{\rm CW}$ to a dynamical, self-dual 2-form
gauge field $b$.
The action becomes
\begin{align}
  \label{Ssixb}
  S_{\mathrm{aE}b}[\phi,a,b]=\int_{\M}\lrC{
-\frac{1}{2}|d\phi|^2
-\frac{1}{2}|da|^2
-\frac{1}{8\pi}
\Big|db-\frac{M}{2\pi}\omega_3(a)\Big|^2
+\frac{N}{48\pi^3}\phi(da)^3
+\frac{M}{8\pi^2}b\wedge (da)^2} \ ,
\end{align}
with $\omega_3(a)=a\wedge da$ being the CS 3-form.
$b$ is normalized as $\int db\in 2\pi\bZ$ and $M\in\bZ$.
The self-duality condition for $b$ is given by
\begin{align}
  db-\frac{M}{2\pi}\omega_3(a)
={}\star\left(db-\frac{M}{2\pi}\omega_3(a)\right) \ .
\label{sd:b}
\end{align}
As a consistency check, we note that the EoM reads
\begin{align}
  \frac{1}{4\pi}
d\star\left(db-\frac{M}{2\pi}\omega_3(a)\right)
+\frac{1}{8\pi^2}(da)^2=0 \ ,\nn
\end{align}
on which imposing (\ref{sd:b}) leads to
the Bianchi identity for the field strength
$db-(M/2\pi)\,\omega_3(a)$.
This field strength is left unchanged under the $U(1)^{[0]}$
gauge transformation if $b$ makes a GS-like transformation as
\begin{align}
  \delta^{[0]}b=\frac{M}{2\pi}\lambda_0 da \ .\nn
\end{align}
The GSWS term 
\begin{align}
  \frac{M}{8\pi^2}b\wedge (da)^2 \nn
\end{align}
is utilized for canceling reducible anomalies in 6d
\cite{Green:1984bx,Sagnotti:1992qw}.

As an example for the GSWS mechanism, we consider
$n$ right-handed fermions of $U(1)^{[0]}$ charge $p$.
They lead to the one-loop box anomaly
\begin{align}
  2\pi I_6=
-n\,p^4
\frac{1}{4!(2\pi)^3}\lambda_0\, (da)^3\ ,
\label{I6}
\end{align}
which descends from the anomaly polynomial
\begin{align}
  I_{8}=-n\,p^4\frac{1}{4!(2\pi)^4}(da)^4 \ .\nn
\end{align}
By imposing 
\begin{align}
  2\pi I_6+\delta^{[0]}S=0 \ ,\nn
\end{align}
with
\begin{align}
  \delta^{[0]}S=\frac{M}{8\pi^2}\delta^{[0]}b\wedge (da)^2
=\frac{M^2}{2(2\pi)^3}\lambda_0\, (da)^3 \ ,\nn
\end{align}
we find
\begin{align}
  M=\left(\frac{n\,p^4}{12}\right)^{1/2}\ .
\label{Mq2}
\end{align}
This is consistent if the RHS is integer.

The axion electrodynamics coupled with the right-handed fermions
is obtained by the Peccei-Quinn(PQ) mechanics \cite{Peccei:1977hh}. 
To see this,
consider the model
\begin{align}
  \begin{array}{ccc}
 & U(1)_L\times U(1)_R    \\
n_L\times\psi_L & (p,0) \\
(n_L+n)\times\psi_R & (0,p) \\
\Phi& (p,-p)\\
b & (0,0)
\label{PQ1}
  \end{array}
\end{align}
$\psi_{L,R}$ are left- and right-handed chiral fermions, respectively. 
The vector-like subgroup of $U(1)_L\times U(1)_R$ is dynamically gauged 
with the dynamical gauge field given by $a$.
We note that the one-loop $U(1)^4$ anomaly from $\psi_L$ and $\psi_R$
is given by (\ref{I6}). This is canceled by the GSWS mechanics
by setting $M$ equal to (\ref{Mq2}).
$\Phi$ is a complex scalar field that is neutral under $U(1)^{[0]}$.
They are coupled with the Yukawa coupling as
\begin{align}
  \cL_{\rm Y}=\Phi\,\psi_L^{i\dagger}\psi_{R i}+{\rm c.c.} \ .\nn
\end{align}
with $i=1,2,\cdots,n_L$. The rest of $\psi_{R i}$ with 
$i=n_L+1,\cdots,n$
has no Yukawa coupling with $\psi_L$.

The PQ mechanism occurs by turning on the vev of $\Phi$:
\begin{align}
  \langle\Phi\rangle = v \ .\nn
\end{align}
This breaks $U(1)_A$ while the $U(1)^{[0]}$ gauge symmetry is left unbroken.
The fluctuations about this vacuum are given by
\begin{align}
  \Phi=(v+\rho)e^{i\phi}\ .\nn
\end{align}
$\rho$ is massive and thus can be integrated out. 
The Yukawa term becomes
\begin{align}
  \cL_{\rm Y}=ve^{i\phi}\psi_L^{\dagger}\psi_R+\cdots \ .\nn
\end{align}
The $U(1)_A$ transformation
\begin{align}
  \psi_L \to e^{ip\alpha}\psi_L \ ,~~~
  \psi_R \to e^{-ip\alpha}\psi_R \ ,\nn
\end{align}
with
\begin{align}
  \phi-2p\alpha=0 \nn
\end{align}
removes $\phi$ from the Yukawa coupling.
Then, $n_L$ pairs of the chiral fermions $\psi_L$ and $\psi_R$ gain
a non-vanishing, $\phi$-independent mass term and thus
can be integrated out.
The $U(1)_A$ rotation gives rise to the axion coupling via $U(1)_A$ 
anomaly
\begin{align}
  2\times n_L\,p^4\alpha\frac{(da)^3}{3!(2\pi)^3}
=n_L\,p^3\phi\frac{(da)^3}{3!(2\pi)^3} \ .\nn
\end{align}
This allows us to identify
\begin{align}
  N=n_L\,p^3 \ .
\label{NnL}
\end{align}

Our main focus in this paper is on a model defined by
gauging the 3-form CW symmetry.
This is done by promoting the CW background gauge field
$D_4^{\mathrm{CW}}$  to a dynamical 4-form gauge field $u$,
which couples minimally with the corresponding CW current as
\begin{align}
  K u\wedge j_{\rm{CW}}^{[3]}
= \frac{K}{2\pi} u\wedge da \ .\nn
\end{align}
Here, $K\in\bZ$ because of the normalization $\int du\in 2\pi\bZ$.
Then, this model is anomaly free by requiring
that $u$ make a GS-like transformation under $U(1)^{[0]}$ as
\begin{align}
 \delta^{[0]}u=- \frac{M^2}{8\pi^2K}\,\lambda_0 \wedge (da)^2 
\ .\nn
\end{align}
This implies that the gauge invariant field strength of $u$ is 
given by
\begin{equation}
  i_5 = du +\frac{M^2}{8\pi^2K}\,a\wedge (da)^2
= du +\frac{M^2}{8\pi^2K}\,\omega_5(a) \ .
\label{def:i5}
\end{equation}

To summarize, the gauge symmetry and the gauge transformation
of the present model reads
\begin{itemize}
  \item 
$U(1)^{[0]}$

\begin{align}
  \delta^{[0]}a=d\lambda_0 \ ,~~
  \delta^{[0]}b=\frac{M}{2\pi}\lambda_0 da \ ,~~
  \delta^{[0]}u=-\frac{M^2}{8\pi^2K}\lambda_0 (da)^2 \ .
\label{u(1)0}
\end{align}

\item
$U(1)^{[1]}$
\begin{align}
  \delta^{[1]}b=d\lambda_1 \ .\nn
\end{align}

\item
$U(1)^{[3]}$
\begin{align}
  \delta^{[3]}u=d\lambda_3 \ .\nn
\end{align}

\end{itemize}
The consistency of (\ref{u(1)0}) with 
the quantization condition $\int du\in 2\pi\bZ$ requires 
\begin{align}
  \frac{M^2}{8\pi^2 K}(2\pi)^3 \in 2\pi \bZ \ .\nn
\end{align}
This is equivalent to stating
that there exists an integer $r_1$ such that
\begin{align}
  M^2=2Kr_1\ .\nn
\end{align}
This shows that $M$ is even. 
Let $m_1=\mbox{gcd}(K,r_1)$. Then,
\begin{align}
  K=m_1\wt K \ ,~r_1=m_1\wt r_1 \ ,
\label{Kr1gcd}
\end{align}
with $\wt K, \wt r_1\in\bZ$ being relatively prime. 
These satisfy
\begin{align}
  M^2=2m_1^2\wt K\,\wt r_1 \ .\nn
\end{align}
This implies that there exists an integer $L_1$ such that
\begin{align}
  2\wt K\,\wt r_1=L_1^2 \ ,
\label{KrL1sq}
\end{align}
which requires that $L_1$ be even. It follows that $M$ is written in terms
of $m_1$ and $L_1$ as
\begin{align}
  M=m_1L_1 \ .
\label{MmL1}
\end{align}

\section{EoM-based current and global symmetry}

Start with the action
\begin{align}
  S
=-\int_{\M}\lrC{\frac{1}{2}|d\phi|^2+\frac{1}{2}|da|^2
+\frac{1}{8\pi}|db-\frac{M}{2\pi}\omega_3(a)|^2
+\frac{1}{2}|du+\frac{M^2}{8\pi^2K}\omega_5(a)|^2}
+S_{\mathrm{CS}}\ .
\label{6daction}
\end{align}
Here
\begin{align}
S_{\mathrm{CS}}
&=\int_{\cM_6}\left(\frac{N}{48\pi^3}\phi(da)^3
+\frac{M}{8\pi^2}(da)^2\wedge b
+\frac{K}{2\pi}\,da\wedge u
\right)
\nn\\
&=\int_{\Omega_{\cM_6}}
\left(\frac{N}{48\pi^3}d\phi\wedge (da)^3
+\frac{M}{8\pi^2}(da)^2\wedge db
+\frac{K}{2\pi}\,da\wedge du
\right)
\ ,\nn
\end{align}
where $\Omega_{\cM_6}$ is a seven-manifold with $\partial\Omega_{\cM_6}=\cM_6$.
The EoM for $\phi$ reads
\begin{align}
&  d^\star d\phi+\frac{N}{48\pi^3}\,(da)^3=0  \ .\nn
\end{align}
This defines the conserved current
\begin{align}
  j_{\mathrm{EoM}}^{[0]}=-\star d\phi-\frac{N}{48\pi^3}\,\omega_5 \ .\nn
\end{align}
As discussed in \cite{Nakajima:2022feg}, this leads to
a 0-form global $\bZ_N$ symmetry, which acts as
\begin{align}
  \delta_{\Lambda_0}\phi=\Lambda_0 \ ,\nn
\end{align}
with $\Lambda_0=2\pi/N$.

The EoM for $u$ reads
\begin{align}
  d\star\left(du+\frac{M^2}{8\pi^2K}\,\omega_5\right)
+\frac{K}{2\pi}\,da=0 \ .
\label{eomu4}
\end{align}
This defines the conserved current
\begin{align}
j_{\mathrm{EoM}}^{[4]}=
-\star\left(du+\frac{M^2}{8\pi^2K}\,\omega_5\right)-\frac{K}{2\pi}\,a 
\ .\nn
\end{align}
Noting that $i_5=du+\frac{M^2}{8\pi^2K}\,\omega_5$ is gauge invariant,
the symmetry generator associated with $j_{\mathrm{EoM}}^{[4]}$ is 
gauge invariant if it generates a 4-form $\bZ_K$ global
symmetry. 
This acts as
\begin{align}
  \delta_{\Lambda_4}u=\Lambda_4 \ ,\nn
\end{align}
with
\begin{align}
  \int \Lambda_4\in\frac{2\pi}{K}\bZ \ .\nn
\end{align}

The EoM for $b$ reads
\begin{align}
  \frac{1}{4\pi}d\star\left(db-\frac{M}{2\pi}\,\omega_3\right)
+\frac{M}{8\pi^2}\,(da)^2=0 \ .
\label{eomb}
\end{align}
The EoM-based conserved current is given by
\begin{align}
j_{\mathrm{EoM}}^{[2]}=
-\frac{1}{4\pi}\star\left(db-\frac{M}{2\pi}\,\omega_3\right)
-\frac{M}{8\pi^2}\,\omega_3
=-\frac{1}{4\pi}db 
\ .\nn
\end{align}
Identification of the global symmetry generated by this current
is tricky because of the self-duality condition of $b$.
One way to do that is to dimensionally reduce the 6d action (\ref{6daction})
to 5d, where $b$ reduces to a $U(1)$ gauge field.
{}For details, see the appendix \ref{DR}.
As discussed there, the resultant action has a 1-form global 
$\bZ_{M}$ symmetry, which is the manifestation of
the 2-form global $\bZ_{M}$ symmetry in 6d after dimensional
reduction.
This acts as
\begin{align}
  \delta_{\Lambda_2}b=\Lambda_2 \ ,\nn
\end{align}
with
\begin{align}
  \int \Lambda_2\in\frac{2\pi}{M}\bZ \ .\nn
\end{align}

The EoM for $a$ reads
\begin{align}
&  d^\star da
-\frac{2M}{16\pi^2}\left[
da\wedge\star\left(db-\frac{M}{2\pi}\omega_3\right)
+d\left\{
a\wedge\star\left(db-\frac{M}{2\pi}\omega_3\right)\right\}
  \right]
\nn\\
+&\frac{M^2}{8\pi^2K}\left[
(da)^2\wedge\star\left(du+\frac{M^2}{8\pi^2K}\omega_5\right)
+2\,d\left\{
\omega_3\wedge\star\left(
du+\frac{M^2}{8\pi^2K}\omega_5\right)
\right\}
\right]
\nn\\
-&\frac{N}{48\pi^3}\,3\,d\phi\wedge(da)^2-\frac{M}{8\pi^2}\,2db\wedge da
-\frac{K}{2\pi}du = 0 \ .
\label{eoma}
\end{align}
This can be written in a gauge invariant form as
\begin{align}
  &d^\star da-\frac{M}{2\pi^2}\,da\wedge \left(db-\frac{M}{2\pi}\omega_3\right)
+\frac{3M^2}{8\pi^2K}\,(da)^2\wedge\star\left(
du+\frac{M^2}{8\pi^2K}\,\omega_5\right)
-\frac{K}{2\pi}\,\left(du+\frac{M^2}{8\pi^2K}\,\omega_5\right)
\nn\\
&-\frac{N}{16\pi^3}\,d\phi\wedge(da)^2=0 \ .
\label{eoma2}
\end{align}
It can be verified that
the LHS of (\ref{eoma2}) is an exact form leading to
an EoM-based conserved current. To see this, we note that (\ref{eomu4})
is integrated as
\begin{align}
  \star\left(du+\frac{M^2}{8\pi^2K}\omega_5\right)
=-\frac{K}{2\pi}\,a+\frac{1}{2\pi}\,dv_0 \ .\nn
\end{align}
Here, $v_0$ is a 0-form field, which is required to transform
as
\begin{align}
\delta^{[0]}v_0=K\lambda_0 \ ,~~
\delta^{[1]}v_0=\delta^{[3]}v_0=0 \ .
\label{gaugetr:v0}
\end{align}
It then follows that
\begin{align}
  du+\frac{M^2}{8\pi^2K}\omega_5
=-\frac{K}{2\pi}\star\left(a-\frac{1}{K}\,dv_0\right) \ .
\label{int:du4}
\end{align}
Inserting this into (\ref{eoma2}) gives
\begin{align}
  d\star da-\frac{M}{2\pi^2}\,da\wedge db
+\frac{3M^2}{16\pi^3K}\,(da)^2\wedge dv_0
-\frac{K}{2\pi}\,du-\frac{N}{16\pi^3}\,d\phi\wedge (da)^2=0 \ .\nn
\end{align}
This can be integrated as
\begin{align}
  j_4\equiv \star da-\frac{M}{2\pi^2}\,da\wedge b
+\frac{3M^2}{16\pi^3K}\,(da)^2 v_0
-\frac{K}{2\pi}\,u-\frac{N}{16\pi^3}\,\phi\,(da)^2
=-\frac{1}{2\pi}dv_3 \ ,
\label{int:eoma}
\end{align}
with $v_3$ being a 3-form gauge field. 
$j_4$ defines an EoM-based conserved current.
It is easy to show that $v_3$ transforms as
\begin{align}
  \delta^{[0]}v_3=0 \ ,~~
  \delta^{[1]}v_3=\frac{M}{\pi}\,\lambda_1\wedge da \ ,~~
  \delta^{[3]}v_3=K\lambda_3 \ .
\label{gaugetr:v3}
\end{align}

As clear from (\ref{gaugetr:v0}) and (\ref{gaugetr:v3}), 
$v_0$ and $v_3$ can be regarded as NG bosons associated with
the spontaneous symmetry breaking for the gauge group $U(1)^{[0]}$
and $U(1)^{[3]}$, respectively. 
This is a manifestation of
the St\"uckelberg mechanism, which gives rise
to a mass for the gauge field $a$ and $u$ via
the coupling $(K/2\pi)\, da\wedge u$.
We normalize
$v_0$ and $v_3$ as
\begin{align}
  \int dv_0 \in 2\pi\bZ \ ,~~~
  \int dv_3 \in 2\pi\bZ \ .\nn
\end{align}
Then, the St\"uckelberg mechanism is equivalent to
Higgsing the gauge groups $U(1)^{[0]}$ and
$U(1)^{[3]}$ by the vev of operators of $K$ units of charges, so that
the gauge groups are broken to $\bZ_K^{[0]}$ and
$\bZ_K^{[3]}$, respectively \cite{Banks:2010zn}.

We can improve $j_{\mathrm{EoM}}^{[1]}$ by adding a total derivative term
to eliminate $v_3$:
\begin{align}
  j_{\mathrm{EoM}}^{\prime [1]}
&=j_{\mathrm{EoM}}^{[1]}+d\Omega_3 
\nn\\
&=\star da-\frac{M}{8\pi^2}\,da\wedge b
-\frac{3M}{4\pi K}\left(
db-\frac{M}{2\pi}\,\omega_3\right)\wedge
\star\left(du+\frac{M^2}{8\pi^2K}\,\omega_5\right)
-\frac{K}{2\pi}\,u-\frac{N}{16\pi^3}\,\phi\,(da)^2 \ ,
\end{align}
where
\begin{align}
  \Omega_3=\frac{3M}{8\pi^2 K}\left(db-\frac{M}{2\pi}\,\omega_3\right)v_0
+\frac{3M}{8\pi^2}\,a\wedge b \ .\nn
\end{align}

We now show that the transformation laws generated by the EoM-based
current are given by
\begin{align}
  \delta_{\Lambda_1}a=\Lambda_1 \ ,~~~
  \delta_{\Lambda_1}b=-\frac{M}{2\pi}\Lambda_1\wedge a \ ,~~~
  \delta_{\Lambda_1}u=\Lambda_1\wedge
\left(
\frac{M^2}{8\pi^2 K}\omega_3-\frac{3M}{4\pi K}db\right) \ .
\label{eomsym}
\end{align}
{}For this purpose,
we gauge $\Lambda_1$ with
$d\Lambda_1\ne 0$.
Using
\begin{align}
  \delta_{\Lambda_1}\left(db-\frac{M}{2\pi}\omega_3\right)
&=-\frac{M}{\pi}d\Lambda_1\wedge a \ ,\nn
\\
  \delta_{\Lambda_1}\left(du+\frac{M^2}{8\pi^2 K}\omega_3\right)
&=-\frac{3M}{4\pi K}d\Lambda_1\wedge 
\left(db-\frac{M}{2\pi}\omega_3\right) \ ,\nn
\end{align}
it is found that
\begin{align}
  \delta_{\Lambda_1}\cL=
d\Lambda_1\wedge
\bigg[
&-\star da
+\frac{M}{4\pi^2}a\wedge\star\left(db-\frac{M}{2\pi}\omega_3\right)
+\frac{3M}{4\pi K}\left(db-\frac{M}{2\pi}\omega_3\right)
\wedge\star
\left(du+\frac{M^2}{8\pi^2 K}\omega_5\right)
\nn\\
&+\frac{N}{16\pi^3}\phi(da)^3
+\frac{M}{4\pi^2}da\wedge b
+\frac{K}{2\pi}u
\bigg]
-\frac{3M}{8\pi^2}\Lambda_1\wedge da\wedge db \ .\nn
\end{align}
Using
\begin{align}
  \Lambda_1\wedge da\wedge db=
d\Lambda_1\wedge\left( a\wedge db+t\cdot d(a\wedge b)\right)
-d\left[
\Lambda_1\wedge\left(
a\wedge db+t\cdot d(a\wedge b)\right)\right] \ ,\nn
\end{align}
with $t$ being any real number, it is found that
\begin{align}
  \delta_{\Lambda_1}\cL
=&
\,d\Lambda_1\wedge
\bigg[
-\star da
+\frac{M}{4\pi^2}a\wedge\star\left(db-\frac{M}{2\pi}\omega_3\right)
+\frac{3M}{4\pi K}\left(db-\frac{M}{2\pi}\omega_3\right)
\wedge\star
\left(du+\frac{M^2}{8\pi^2 K}\omega_5\right)
\nn\\
&\qquad\quad+\frac{N}{16\pi^3}\phi(da)^3
+\frac{K}{2\pi}u
+\frac{M}{4\pi^2}\left(1-\frac{3t}{2}\right)da\wedge b
+\frac{3M}{8\pi^2}\left(-1+t\right)a\wedge db
\bigg]
\nn\\
&+\frac{3M}{8\pi^2}
d\left[
\Lambda_1\wedge\left(
a\wedge db+t\cdot d(a\wedge b)\right)\right]\ .\nn
\end{align}
Imposing the self-duality condition (\ref{sd:b}) reduces it to
\begin{align}
  \delta_{\Lambda_1}\cL
=&
\,d\Lambda_1\wedge
\bigg[
-\star da
+\frac{3M}{4\pi K}\left(db-\frac{M}{2\pi}\omega_3\right)
\wedge\star
\left(du+\frac{M^2}{8\pi^2 K}\omega_5\right)
+\frac{N}{16\pi^3}\phi(da)^3
+\frac{K}{2\pi}u
\nn\\
&
\qquad\quad+\frac{M}{4\pi^2}\left(1-\frac{3t}{2}\right)da\wedge b
+\frac{M}{8\pi^2}\left(-1+3t\right)a\wedge db
\bigg]
\nn\\
&+\frac{3M}{8\pi^2}
d\left[
\Lambda_1\wedge\left(
a\wedge db+t\cdot d(a\wedge b)\right)\right]\ .\nn
\end{align}
Setting $t=1/3$ and restoring $\Lambda_1$ to a global transformation
parameter, we obtain the conservation law for 
$j_{\mathrm{EoM}}^{\prime [1]}$.

Let
\begin{align}
  U^{[1]}(\alpha,\cM_4)=e^{i\alpha\int_{\cM_4} j_{\mathrm{EoM}}^{\prime [1]}} 
\nn
\end{align}
be the symmetry generator defined by
$j_{\mathrm{EoM}}^{\prime [1]}$. 
The transformation parameter $\alpha$ is written in terms of $\Lambda_1$ as
\begin{align}
  \alpha=\int \Lambda_1 \ ,\nn
\end{align}
and constrained from the
gauge invariance of the symmetry generator.
The conserved charge $\int_{\cM_4} j_{\mathrm{EoM}}^{\prime [1]}$ is rewritten
in a gauge invariant form by lifting it to a five-dimensional
manifold. The gauge invariance of $U^{[1]}$ amounts to
requiring that it be independent of how to lift it. We are led to
impose the condition
\begin{align}
  \exp i\alpha\int_{\cM_5}
\left(
-\frac{M}{8\pi^2}da\wedge db-\frac{K}{2\pi}du
-\frac{N}{16\pi^3}d\phi\wedge (da)^2\right)=1 \ ,
\label{gaugeinv:uj4}
\end{align}
for any closed manifold $\cM_5$.
By using
\begin{align}
  \frac{1}{2\pi}\int_{\cM_5}du&=n_1 \ ,~~~(n_1\in\bZ)\nn
\\
  \frac{1}{8\pi^2}\int_{\cM_5}da\wedge db&=\frac{n_2}{2} \ ,~~~(n_2\in\bZ)\nn
\\
  \frac{1}{16\pi^3}\int_{\cM_5}d\phi\wedge (da)^2&=\frac{n_3}{2} \ ,
~~~(n_3\in\bZ)\nn
\end{align}
we obtain
\begin{align}
  \left(
n_1K+\frac{n_2}{2}M+\frac{n_3}{2}N
\right)\alpha\in 2\pi\bZ \ .\nn
\end{align}
Using (\ref{Kr1gcd}) and (\ref{MmL1}), we find
\begin{align}
  n_1K+\frac{n_2}{2}M+\frac{n_3}{2}N
=m_1\left(n_1\wt K+n_2\frac{L_1}{2}\right)+\frac{N}{2}n_3 \ .\nn
\end{align}
Note that $L_1/2\in\bZ$. If there exists an nontrivial integer $q$
such that
\begin{align}
  q={\rm{gcd}}(m_1,\frac{N}{2}) \ ,
\label{qm1N}
\end{align}
the symmetry generator is gauge invariant by setting
\begin{align}
  \alpha\in\frac{2\pi}{q}\bZ \ .\nn
\end{align}
This shows that the EoM-based current leads to the global 1-form
$\bZ_q$ symmetry.
It follows from (\ref{qm1N}) that $m_1$ and $N$ are written as
\begin{align}
  m_1=q\wt m_1 \ ,~~~
  N=q\wt N \ ,
\label{m1Ntilde}
\end{align}
with $\wt m_1,\,\wt N/2\in\bZ$ relatively prime with each other.

As a consistency check for the existence of $\bZ_q^{[1]}$,
we show that (\ref{eomsym}) is consistent with
the normalization condition of $b$ and $u$.
It is found that
\begin{align}
  \delta_{\Lambda_1}\int db&=-\frac{M}{2\pi}\int\Lambda_1\wedge da
\in \frac{2\pi M}{q}\bZ
 \ ,\nn
\\
  \delta_{\Lambda_1}\int du&=\frac{M^2}{8\pi^2 K}\int\Lambda_1\wedge (da)^2
\in \frac{2\pi M^2}{Kq}\bZ
 \ .\nn
\end{align}
Using (\ref{MmL1}) and (\ref{m1Ntilde}) gives 
\begin{align}
  \frac{M}{q}=\wt m_1 L_1 \ ,\nn
\end{align}
guaranteeing that the normalization condition for $b$ 
is consistent with the global $\bZ_q^{[1]}$ symmetry
transformation.
{}Furthermore, it is verified that
\begin{align}
\frac{M^2}{Kq}=\frac{M}{K}\wt m_1 L_1
=\frac{\wt m_1 L_1^2}{\wt K}
=2\wt m_1 \wt r_1 \ .  \nn
\end{align}

\subsection{Dualizing the 4-form gauge field}

It is interesting to rewrite the Lagrangian by dualizing $u$.
The gauge invariant field strength of $u$  is given
by (\ref{def:i5}),
which obeys the Bianchi identity
\begin{align}
  di_5=\frac{M^2}{8\pi^2 K}(da)^3 \ .\nn
\end{align}
Instead of $u$, we regard $i_5$ as an independent variable.
This is achieved by using a Lagrange multiplier $\xi$ that
imposes the Bianchi identity as a constraint. We consider
\begin{align}
  S^\prime&=S
+\frac{1}{2\pi}\int\xi\left(di_5-\frac{M^2}{8\pi^2 K}(da)^3\right)
\nn\\
&=
\int_{\M}\bigg\{-\frac{1}{2}|d\phi|^2-\frac{1}{2}|da|^2
-\frac{1}{8\pi}|db-\frac{M}{2\pi}\omega_3|^2
+\frac{N}{48\pi^3}\phi(da)^3+\frac{M}{8\pi^2}(da)^2 b
\nn\\
&\qquad\qquad
-\frac{1}{2}|i_5|^2
+\frac{1}{2\pi} i_5\wedge\left(d\xi-Ka\right)
-\frac{M^2}{16\pi^3 K}\xi(da)^3
\bigg\}
\ .\nn
\end{align}
Here, $\xi$ is a compact scalar with $\xi\sim\xi+2\pi$.
The EoM of $i_5$ reads
\begin{align}
  {}\star i_5=\frac{1}{2\pi}\left(d\xi-Ka\right) \ .
\label{eom:i5}
\end{align}
The $U(1)^{[0]}$ gauge invariance of $i_5$ requires that
$\xi$ transform under it as
\begin{align}
  \delta^{[0]}\xi=K\lambda_0 \ .
\label{del0xi}
\end{align}
It follows that $\xi$ is identified with an NG boson associated with
the spontaneous breaking of $U(1)^{[0]}$ to $\bZ_K^{[0]}$.
This is consistent with those obtained in the previous section,
where the NG boson associated with the symmetry breaking
$U(1)^{[0]}\to \bZ_K^{[0]}$ is given by $v_0$.

By using (\ref{eom:i5}), the action becomes
\begin{align}
  S^\prime
&=
\int_{\M}\bigg\{-\frac{1}{2}|d\phi|^2-\frac{1}{2}|da|^2
-\frac{1}{8\pi^2}
\left|d\xi-Ka\right|^2
-\frac{1}{8\pi}|db-\frac{M}{2\pi}\omega_3|^2
\nn\\
&\qquad\qquad
+\frac{N}{48\pi^3}\phi(da)^3
-\frac{M^2}{16\pi^3 K}\xi(da)^3
+\frac{M}{8\pi^2}(da)^2 \wedge b
\bigg\} \ .
\label{S:u4dual}
\end{align}
As a consistency check, we see that the axionic coupling
between $\xi$ and $(da)^3$ is gauge invariant by showing that
it is properly normalized:
\begin{align}
  \frac{M^2}{16\pi^2K}
=\frac{1}{48\pi^3}\frac{3M^2}{K}
=\frac{6r_1}{48\pi^3} \ ,\nn
\end{align}
with $r_1\in\bZ$.
{}Furthermore, we note that the anomaly from the GSWS transformation
$\delta^{[0]}b$ is canceled by that from $\delta^{[0]}\xi$
thanks to the axionic coupling.

\section{Gauging the EoM-based global symmetry}

We gauge the EoM-based global symmetry 
by promoting $\lambda_{0,1,2,4}$
to non-closed, gauge transformation functions.\footnote{We use
the terminology ``gauging'' also for defining a local symmetry
whose gauge field is a background field.}
The gauge invariant action can be obtained by activating
the corresponding background gauge fields.

{}For the 0-form $\bZ_N$ symmetry, we consider
\begin{align}
  (A_1,A_0) \ ,~~NA_1=dA_0 \ .\nn
\end{align}
The field strength for $\phi$ is defined
as $d\phi-A_1$.

{}For the 2-form $\bZ_M$ symmetry, we consider
\begin{align}
  (V_3,V_2) \ ,~~MV_3=dV_2 \ .\nn
\end{align}
The field strength for $b$ is given by
$db-V_3$.

{}For the 4-form $\bZ_K$ symmetry, we consider
\begin{align}
  (W_5,W_4) \ ,~~KW_5=dW_4 \ ,\nn
\end{align}
leading to the field strength $du-W_5$ for $u$.

We discuss the gauging of the 1-form $\bZ_q$ symmetry
in some detail.
We turn on the associated background gauge field
\begin{align}
  (B_2,B_1) \ ,~~qB_2=dB_1 \ .\nn
\end{align}
It follows from (\ref{eomsym}) that
the field strength for $a$ is given simply by
$da-B_2$.
{}For the purpose of defining the field strength of $b$
and $u$ for the $\bZ_q$ gauge symmetry, we first modify the
transformation law for $b$ in (\ref{eomsym}) so that
the RHS becomes $\bZ_q$-invariant:
\begin{align}
  \delta_{\Lambda_1}b=-\frac{M}{2\pi}\Lambda_1\wedge 
\left(a-\frac{1}{q}B_1\right) \ .\nn
\end{align}
By noting $\Lambda_1=\frac{1}{q}\,\delta_{\Lambda_1}B_1$,
this is rewritten as
\begin{align}
  \delta_{\Lambda_1}\ol b=0 \ , ~~~
  \ol b\equiv b+\frac{M}{2\pi q}B_1\wedge 
\left(a-\frac{1}{q}B_1\right) \ .\nn
\end{align}
The field strength for $b$ is defined naturally as
\begin{align}
d\ol b
=db-\frac{M}{2\pi q}da\wedge B_1
+\frac{M}{2\pi}a\wedge B_2
\ .\nn
\end{align}
We also modify the transformation law of $u$ in (\ref{eomsym}) 
so that the RHS becomes $\bZ_q$-invariant:
\begin{align}
  \delta_{\Lambda_1}u=\Lambda_1\wedge
\left(
\frac{M^2}{8\pi^2K}\,\omega_3(a-\frac{1}{q}B_1)
-\frac{3M}{4\pi K}
\left(
db-\frac{M}{2\pi q}da\wedge B_1+\frac{M}{2\pi}a\wedge B_2
\right)
\right)\ .\nn
\end{align}
It then follows that
\begin{align}
  \delta_{\Lambda_1}
\ol u=0 \ ,~~~
\ol u\equiv
u-\frac{1}{q}B_1\wedge
\left(
\frac{M^2}{8\pi^2K}\,\omega_3(a-\frac{1}{q}B_1)
-\frac{3M}{4\pi K}
d\ol b
\right) \ .\nn
\end{align}
The field strength for $u$ reads
\begin{align}
d\ol u
=du
-B_2\wedge
\left(
\frac{M^2}{8\pi^2K}\,\omega_3(a-\frac{1}{q}B_1)
-\frac{3M}{4\pi K}
d\ol b
\right)
+\frac{M^2}{8\pi^2Kq}B_1\wedge
(da-B_2)^2 \ .\nn
\end{align}

Now we examine 
the combined gauge transformation
of $U(1)^{[0]}$ and $\bZ_q^{[1]}$ for $b$ and $u$ to define
the corresponding field strengths.
We first assume that it is given by the linear combonation
of $\delta^{[0]}$ and $\delta_{\Lambda_1}$ with the gauge field $a$ 
appearing in $\delta^{[0]}$ replaced by $a-\frac{1}{q}B_1$.
Then, it follows that
\begin{align}
  b\overset{?}{\to} b+\frac{M}{2\pi}\lambda_0(da-B_2)
-\frac{M}{2\pi}\Lambda_1\wedge\left(a-\frac{1}{q}B_1\right)
\ .\nn
\end{align}
Then, $\ol b$ transforms as
\begin{align}
  \ol b&\overset{?}{\to} \ol b+\frac{M}{2\pi}\lambda_0(da-B_2)
-\frac{M}{2\pi q}d\lambda_0\wedge 
\left(B_1+q\Lambda_1 \right)
\nn\\
&=
\ol b+\frac{M}{2\pi}\lambda_0(da-B_2)
-d\left(\frac{M}{2\pi q}\lambda_0
\left(B_1+q\Lambda_1\right)\right)
+\frac{M}{2\pi}\lambda_0
\left(B_2+d\Lambda_1 \right)
\ .\nn
\end{align}
The 4th term 
makes it impossible to define a gauge invariant field strength
even using a GS-like transformation. 
This problem is resolved 
by modifying the gauge transformation
as
\begin{align}
  b\to b+\frac{M}{2\pi}\lambda_0(da-B_2)
-\frac{M}{2\pi}\lambda_0
\left(B_2+d\Lambda_1 \right)
-\frac{M}{2\pi}\Lambda_1\wedge\left(a-\frac{1}{q}B_1\right)
\ ,\nn
\end{align}
from which we obtain
\begin{align}
  \ol b\to
\ol b+\frac{M}{2\pi}\lambda_0(da-B_2)
-d\left(\frac{M}{2\pi q}\lambda_0
\left(B_1+q\Lambda_1\right)\right)
\ .\nn
\end{align}
We are thus led to define the field strength of $b$ for the
combined gauge transformation:
\begin{align}
D b=  d\ol b-\frac{M}{2\pi}\omega_3(a-\frac{1}{q}B_1) \ .\nn
\end{align}
It is natural to work with $\ol b$ rather than $b$
as an independent, dynamical field.
To see this, we note that the combined gauge transformation of $db$ 
is given by
\begin{align}
  db\to db+\frac{M}{2\pi}d\lambda_0\wedge (da-B_2)
-\frac{M}{2\pi}d\lambda_0\wedge
\left(B_2+d\Lambda_1 \right)
-\frac{M}{2\pi}d\left(\Lambda_1\wedge\left(a-\frac{1}{q}B_1\right)\right)
\ .\nn
\end{align}
This shows that the gauge transformation 
spoils the normalization condition $\int db\in 2\pi\bZ$.
On the contrary, by using
\begin{align}
  d\ol b\to
d\ol b+\frac{M}{2\pi}d\lambda_0\wedge (da-B_2)
\ ,\nn
\end{align}
the gauge transformation acts on $\int d\ol b$ as a
$2\pi\frac{M}{q}\bZ$ shift.
This is consistent because $\int d\ol b$ is not quantized.

The gauge transformation and the field strength of $u$
for the combined gauge transformation can be derived in the same
manner. 
We assume that 
the gauge transformation should be modified from
the linear combination $(\delta^{[0]}+\delta_{\Lambda_1})u$
by adding an additional transformation:
\begin{align}
  u\to
u
-\frac{M^2}{8\pi^2 K}\lambda_0(da-B_2)^2
+\Lambda_1\wedge
\left(
\frac{M^2}{8\pi^2K}\,\omega_3(a-\frac{1}{q}B_1)
-\frac{3M}{4\pi K}
d\ol b
\right)
+\delta^\prime u \ ,\nn
\end{align}
from which
\begin{align}
  \ol u&\to 
\ol u
-\frac{M^2}{8\pi^2 K}\lambda_0(da-B_2)^2
-d\left(\frac{M^2}{4\pi^2 Kq}\lambda_0\left(B_1+q\Lambda_1\right)
\wedge(da-B_2)
\right)
\nn\\
&\qquad~+
\frac{M^2}{4\pi^2 K}\lambda_0
\left(B_2+d\Lambda_1\right)
\wedge(da-B_2)+\delta^\prime u \ .\nn
\end{align}
We set
\begin{align}
  \delta^\prime u=-\frac{M^2}{4\pi^2 K}\lambda_0
\left(B_2+d\Lambda_1\right)
\wedge(da-B_2) \ .\nn
\end{align}
It follows that the field strength for $u$ reads
\begin{align}
D u=d\ol u
+\frac{M^2}{8\pi^2 K}\omega_5(a-\frac{1}{q}B_1)
\ .\nn
\end{align}
As in the case for $b$, we should treat $\ol u$ as an
independent, dynamical field rather than $u$. 

We now derive the action in the presence of the background gauge 
fields. This is obtained
by replacing the derivative of the dynamical
fields in (\ref{6daction}) with the field strengths
we have constructed:
\begin{align}
  S
=-\int_{\M}\left(\frac{1}{2}|d\phi-A_1|^2
+\frac{1}{2}|da-B_2|^2
+\frac{1}{8\pi}|D b-V_3|^2
+\frac{1}{2}\left|D u-W_5\right|^2\right)
+S_{\mathrm{CS}}\ ,
\end{align}
where
\begin{align}
S_{\mathrm{CS}}
&
=\int_{\Omega_{\cM_6}}
\left(\frac{N}{48\pi^3}(d\phi-A_1)\wedge (da-B_2)^3
+\frac{M}{8\pi^2}(da-B_2)^2\wedge (D b-V_3)
+\frac{K}{2\pi}\,(da-B_2)\wedge (D u-W_5)
\right)
\nn\\
&
=\int_{\Omega_{\cM_6}}
\left(\frac{N}{48\pi^3}(d\phi-A_1)\wedge (da-B_2)^3
+\frac{M}{8\pi^2}(da-B_2)^2\wedge (d\ol b-V_3)
+\frac{K}{2\pi}\,(da-B_2)\wedge (d\ol u-W_5)
\right)
\ .
\label{wCS}
\end{align}
As a consistency check, we note that the CS action
is left unchanged manifestly under the small gauge transformations.
{}Furthermore, the integrand of the CS action is a closed 7-form 
so that
it is regarded as a 6d action.

\subsection{Operator-valued ambiguity}

We now examine if the resultant CS action (\ref{wCS}) suffers
an operator-valued ambiguity, which is a quantum 
inconsistency stating that the quantum parts of the CS action
written in terms of the dynamical fields depend
on how to lift them to a seven-manifold.
It is found that there arise the operator-valued
ambiguities as far as the background gauge field $(B_2,B_1)$
is turned on. To see this, focus on the term
\begin{align}
  \int_{\Omega_{\cM_6}}\frac{K}{2\pi}da\wedge d\ol u \ .\nn
\end{align}
This depends on the choice of the seven-manifold $\Omega_{\cM_6}$
even if $K$ is an integer because the integral $\int d\ol b$ over
any closed 3-cycle is not quantized
in the presence of $(B_2,B_1)$.
This is a manifestation of the quantum anomaly of the EoM-based $\bZ_q^{[1]}$
symmetry via the operator-valued ambiguity.
This result is consistent with those obtained in
\cite{Apruzzi:2020zot}.\footnote{See also \cite{BenettiGenolini:2020doj} 
for a related work.}
This paper discusses a discrete 1-form symmetry in $\cN=(1,0)$ 6d 
theories coupled with tensor and vector multiplets. It is shown 
that turning on the background gauge field associated with a 1-form 
symmetry spoils the Dirac quantization condition for BPS string
charges, which gives rise to massive excitation modes that break the 1-form
symmetry explicitly. 

In order to remove the operator-valued ambiguity, we have to
turn off $(B_2,B_1)$. Then, the dynamical field $\ol b$ and $\ol u$
reduces to $b$ and $u$, respectively. The resultant 6d model
has no operator-valued ambiguity, because the flux integral of $b$ and $u$ 
over any closed cycles is quantized.

\section{Discussions}

As discussed in \cite{Nakajima:2022feg}, the 6d axion-electrodynamics
has rich higher-group structure that is manifested as
GS-like transformation laws for the background gauge fields
associated with the CW symmetries. 
A key role is played by the background gauge field for 
a discrete 1-form symmetry, $(B_2,B_1)$, as clear from the
gauge invariant field strengths given in (\ref{FS}).
In this paper, we have investigated the model that
is obtained by promoting the CW background gauge field 
$B_2^{\mathrm{CW}}$ and $D_4^{\mathrm{CW}}$ to the dynamical
field $b$ and $u$, respectively.
It is shown that the background gauge field for the EoM-based
$\bZ_q^{[1]}$ symmetry, which is denoted by
$(B_2,B_1)$ as well, must be turned off because it would
cause an operator-valued ambiguity.
We examine whether there exists a nontrivial higher-group structure
even in the absence of $(B_2,B_1)$.

The CW currents in this model are written in terms of the wedge products of
the gauge invariant closed forms
\begin{align}
  du+\frac{M}{4\pi K}da\wedge db \ ,~~da\ ,~~d\phi \ .\nn
\end{align}
The gauge invariance of the CW 5-form is guaranteed because
\begin{align}
  du+\frac{M}{4\pi K}da\wedge db
=
du+\frac{M^2}{8\pi^2 K}\omega_5(a)+
\frac{M}{4\pi K}da\wedge 
\left(db-\frac{M}{2\pi}\omega_3(a)
\right) \ .\nn
\end{align}
We first compute the integral of the 5-form 
over a closed 5-cycle.
Using (\ref{Kr1gcd}) and (\ref{MmL1}) gives
\begin{align}
  \frac{M}{2K}=\frac{L_1}{2\wt K}\ .\nn
\end{align}
Set
\begin{align}
  s=\rm{gcd}(\frac{L_1}{2},\wt K) \ ,\nn
\end{align}
or equivalently
\begin{align}
  \frac{L_1}{2}=sL^\prime\ , ~~
\wt K=sK^\prime \ ,\nn
\end{align}
with $L^\prime$ and $K^\prime$ relatively prime.
It is found from these and (\ref{KrL1sq}) that
there exists an odd integer $s^\prime$ such that
\begin{align}
  2s=s^\prime K^\prime \ ,\nn
\end{align}
with $K^\prime$ even.
It then follows that
\begin{align}
  \int\left(du+\frac{M}{4\pi K}da\wedge db\right) \in 
\frac{2\pi}{K^\prime}\bZ \ .\nn
\end{align}
It is thus natural to normalize 
the CW 5-form as
\begin{align}
  K^\prime\left(du+\frac{M}{4\pi K}da\wedge db\right)
=
K^\prime du+\frac{L^\prime}{2\pi}da\wedge db \ .\nn
\end{align}

We consider the CW 6-form current
\begin{align}
  d\phi\wedge\left(
K^\prime du+\frac{L^\prime}{2\pi}da\wedge db\right)\ .\nn
\end{align}
Let $\vartheta_0^{\rm{CW}}$ be the CW 0-form background gauge field
coupled with this current
with the normalization condition given by
$\int d\vartheta_0^{\rm{CW}}\in 2\pi\bZ$. 
We propose that 
the interaction term in the presence of 
$(A_1,A_0)$, $(V_3,V_2)$ and $(W_5,W_4)$ 
takes the form
\begin{align}
  S_{\mathrm{CW}}=
\frac{2m_1sQ}{4\pi^2}\int_{\cM_6}
&\vartheta_0^{\mathrm{CW}}\left(d\phi-A_1\right)\wedge
\left[
K^\prime (du-W_5)
+
\frac{L^\prime}{2\pi}
da\wedge(db-V_3)
\right] \ ,
\nn
\end{align}
with $Q\in\bZ$.
It is found that the operator-valued ambiguity comes from
the term
\begin{align}
-\frac{2m_1sQ}{4\pi^2}\int_{\cM_6}
  \vartheta_0^{\mathrm{CW}}A_1\wedge
\left(
K^\prime du
+
\frac{L^\prime}{2\pi}
da\wedge (db-V_3)
\right)\ .
\nn
\end{align}
This is canceled by adding
the local counterterm
\begin{align}
\Delta S
=\frac{2m_1s}{2\pi}\int_{\cM_6}
\left(\eta_1^{\rm{CW}}+\frac{Q}{2\pi}\vartheta_0^{\mathrm{CW}}A_1\right)
\wedge
\left(
K^\prime (du-W_5)
+
\frac{L^\prime}{2\pi}
da\wedge (db-V_3)
\right)
\ .\nn
\end{align}
Here, $\eta_1^{\mathrm{CW}}$ is 
the background gauge field associated with the CW current
$K^\prime du+\frac{L^\prime}{2\pi}da\wedge db$
with $\int d\eta_1^{\rm{CW}}\in 2\pi\bZ$.
It is easy to see that there appears no operator-valued ambiguity
in $S_{\rm{CW}}+\Delta S$.
The gauge invariance of $\Delta S$ induces a GS-like transformation for the
1-form CW gauge field $\eta_1^{\rm{CW}}$ that is
charactorized by the integer $Q$.

So far, we have considered the cases with a nontrivial $K\in\bZ$.
When restricting to $K=1$, we have $q=1$ so that $\bZ_{q}^{[1]}$
symmetry is trivial.
It is seen that this case admits much richer higher-group
structure. 
To  explore it in more detail deserves a further study.

It is found recently that the 4d axion electrodynamics admits
non-invertible higher-form symmetries \cite{Choi:2022fgx,Yokokura:2022alv}.
A study of non-invertible symmetry in the present model and the 
higher-group structure underlying it would be rather interesting.

\appendix

\section{Dimensional reduction to 5d}
\label{DR}

Start with part of the action (\ref{6daction}):
\begin{align}
  &\hat{S}=
\nonumber\\
-&\int_{\M}\lrC{\frac{1}{2}|\hd\hat{\phi}|^2+\frac{1}{2}|\hd\ha|^2
+\frac{1}{8\pi}|\hd\hb-\frac{M}{2\pi}\hat{\omega}_3|^2
+\frac{1}{2}\left|\hd\hu+\frac{M^2}{8\pi^2K}\hat{\omega}_5\right|^2
-\frac{N}{48\pi^3}\hat{\phi}(\hd\ha)^3-\frac{M}{8\pi^2}(\hd\ha)^2 \hb
-\frac{K}{2\pi}\,\hd\ha\, \hu} \ .\nn
\end{align}
The fields with a hat are defined in 6d before dimensional reduction.
Let
\begin{align}
  M,N&=0,1,\cdots,5 \ ,~~~(\mbox{6d spacetime indices})
\nn\\
  \mu,\nu&=0,1,\cdots,4 \ ,~~~(\mbox{5d spacetime indices})\nn
\end{align}
with $X_5\sim X_5+1$. The space-time metric is given by
\begin{align}
  \eta_{MN}={\rm{diag}}(-+++++) \ .\nn
\end{align}

Define the 5d gauge fields as
\begin{align}
  \hb=b_2+b_1\wedge dX_5 \ ,
~~~
\ha=a+\zeta dX_5 \ ,
~~~
\hu=u+u_3\wedge dX_5 \ .\nn
\end{align}
This choice is natural because the fluxes of the resultant 5d 
gauge fields are normalized as $2\pi\bZ$.
Then,
\begin{align}
  \hd\hb-\frac{M}{2\pi}\hat{\omega}_3
=db_2-\frac{M}{2\pi}\omega_3
+\left(db_1-\frac{M}{2\pi}\left(\zeta da-d\zeta\wedge a\right)\right)\wedge dX_5 
\equiv G_3+\cF_2\wedge dX_5
\ .\nn
\end{align}
Let $\hst$ and $\star$ be the Hodge star operation in 6d and 5d, respectively.
Using the formulae
\begin{align}
  \hst\left(
dX^{\mu_1}\wedge\cdots\wedge dX^{\mu_p}\right)
&=
  \star\left(
dX^{\mu_1}\wedge\cdots\wedge dX^{\mu_p}\right)\wedge dX_5 \ ,~~~
\nn\\
  \hst\left(
dX^{\mu_1}\wedge\cdots\wedge dX^{\mu_q}\wedge dX_5\right)
&=
(-)^{q+1}  \star\left(
dX^{\mu_1}\wedge\cdots\wedge dX^{\mu_q}\right) \ ,~~~\nn
\end{align}
we obtain
\begin{align}
  \hst\left(\hd\hb-\frac{M}{2\pi}\hat{\omega}_3\right)
=\star G_3\wedge dX_5-\star\cF_2 \ .\nn
\end{align}
The self-duality condition for $b$ is rewritten as
\begin{align}
  \star G_3=\cF_2 \ , ~~~ \star\cF_2=-G_3 \ .
\label{asd:5d}
\end{align}

The field strength of $\hu_4$ is reduced as
\begin{align}
\hat{i}_5=  \hd\hu_4+\frac{M^2}{8\pi^2K}\hat{\omega}_5
=
i_5+\cF_4\wedge dX_5 \ ,\nn
\end{align}
where
\begin{align}
i_5=du_4+\frac{M^2}{8\pi^2K}\omega_5\ ,
~~~
\cF_4=
du_3+\frac{M^2}{8\pi^2K}
\left(\zeta(da)^2-2d\zeta\wedge\omega_3\right)
\ .\nn
\end{align}
Then,
\begin{align}
  \hst\left(  \hd\hu_4+\frac{M^2}{8\pi^2K}\hat{\omega}_5\right)
=
\left(\star i_5\right)\wedge dX_5-\star\cF_4 \ .\nn
\end{align}

The 0-form gauge symmetry transformation acts on $\ha$ as
\begin{align}
  \hat{\delta}^{[0]}\ha=\hat d\hat{\lambda}_0 \ .
\label{1f-gtr}
\end{align}
Reducing $\hat{\lambda}_0$ to 5d leads to
\begin{align}
  \delta^{[0]} a=d\lambda_0 \ ,\nn
\end{align}
with $\zeta$ left unchanged. $\zeta$ is a compact boson of period $2\pi$.
This is seen from the 0-form gauge transformation with 
$\hat{\lambda}_0=2\pi X_5$.

The 1-form gauge symmetry transformation acts on $\hb$ as
\begin{align}
  \hat{\delta}^{[1]}\hb=\hd\hat{\lambda}_1 \ .\nn
\end{align}
Reducing $\hat{\lambda}_1$ to 5d as
\begin{align}
  \hat{\lambda}_1=\lambda_1+\lambda_0\,dX_5 \ ,\nn
\end{align}
it acts as
\begin{align}
  \delta^{[1]} b_2=d\lambda_1 \ ,~~~
  \delta^{[1]} b_1=d\lambda_0 \ .\nn
\end{align}

The 3-form gauge symmetry transformation acts on $\hu_4$ as
\begin{align}
  \hat{\delta}^{[3]}\hu_4=\hd\hat{\lambda}_3 \ .\nn
\end{align}
Reducing $\hat{\lambda}_3$ to 5d as
\begin{align}
  \hat{\lambda}_3=\lambda_3+\lambda_2\wedge dX_5 \ ,\nn
\end{align}
it acts as
\begin{align}
  \delta^{[3]} u_4=d\lambda_3 \ ,~~~
  \delta^{[3]} u_3=d\lambda_2 \ .\nn
\end{align}

The GS laws for $\hb$ and $\hu_4$ under $U(1)^{[0]}$ are given 
in (\ref{u(1)0}),
which is reduced to 5d as
\begin{align}
\delta^{[0]} b_2=\frac{M}{2\pi}
\lambda_0 da\ ,~~~
\delta^{[0]} b_1=\frac{M}{2\pi}
\lambda_0 d\zeta \ ,\nn
\end{align}
and
\begin{align}
  \delta^{[0]}u_4=-\frac{M^2}{8\pi^2K}\lambda_0(da)^2 \ ,~~
  \delta^{[0]}u_3=-\frac{M^2}{4\pi^2K}\lambda_0 d\zeta\wedge da \ ,\nn
\end{align}
respectively.
Note that these are consistent with the normalization condition
for the 5d gauge fields.

Upon dimensional reduction to 5d, the action becomes
\begin{align}
  S=\int\bigg[
&-\frac{1}{8\pi}\left|G_3\right|^2
-\frac{1}{8\pi}\left|\cF_2\right|^2
+\frac{M}{4\pi^2}b_2\wedge d\zeta\wedge da
+\frac{M}{8\pi^2}b_1\wedge (da)^2
\nn\\
&-\frac{1}{2}\left|i_5\right|^2
-\frac{1}{2}\left|\cF_4\right|^2
+\frac{K}{2\pi}u_4\wedge d\zeta
+\frac{K}{2\pi}u_3\wedge da
-\frac{1}{2}\left|d\phi\right|^2
+\frac{N}{16\pi^3}\phi\, d\zeta\wedge(da)^2
\nn\\
&-\frac{1}{2}|da|^2-\frac{1}{2}|d\zeta|^2
\bigg] \ .\nn
\end{align}
Here, we used $\int dX_5=1$. As a check, the equations of motion
for $b_2$ and $b_1$ read
\begin{align}
  d\star G_3 = -\frac{M}{\pi}\,d\zeta\wedge da \ ,~~~
  d\star \cF_2 = \frac{M}{2\pi}\,(da)^2 \ .
\label{eom:balpha}
\end{align}
These are consistent with the self-duality consition (\ref{asd:5d}).

So far, we regard $b_2$ and $b_1$ as independent variables and
impose the self-duality condition (\ref{asd:5d}) on the equations
of motion. Instead, we attempt to reformulate the action so that $G_3$ 
becomes an independent variable rather than $b_2$. 
{}For this purpose, we note that 
the third term in the 5d action is rewritten as
\begin{align}
  \frac{M}{4\pi^2}b_2\wedge d\zeta\wedge da
=
\Delta\cL
+\frac{1}{4\pi}G_3\wedge\cF_2
-\frac{M^2}{16\pi^3}\zeta\,\omega_5+\frac{M}{8\pi^2}b_1\wedge(da)^2 \ ,\nn
\end{align}
where
\begin{align}
  \Delta\cL=-\frac{1}{4\pi}d\left[
b_2\wedge\cF_2-\frac{M}
{2\pi}
b_1\wedge\omega_3
\right]\ .\nn
\end{align}
Now we define
\begin{align}
  \bar{S}&=S-\int \Delta\cL
\nn\\
=\int\bigg[
&-\frac{1}{8\pi}\left|G_3\right|^2
-\frac{1}{8\pi}\left|\cF_2\right|^2
+\frac{1}{4\pi}G_3\wedge\cF_2
+\frac{M}{4\pi^2}b_1\wedge (da)^2
-\frac{M^2}{16\pi^3}\zeta\,\omega_5
\nn\\
&-\frac{1}{2}\left|i_5\right|^2
-\frac{1}{2}\left|\cF_4\right|^2
+\frac{K}{2\pi}u_4\wedge d\zeta
+\frac{K}{2\pi}u_3\wedge da
-\frac{1}{2}\left|d\phi\right|^2
+\frac{N}{16\pi^3}\phi\, d\zeta\wedge(da)^2
\nn\\
&-\frac{1}{2}|da|^2-\frac{1}{2}|d\zeta|^2
\bigg] \ .\nn
\end{align}
This technique is utilized in \cite{Bonetti:2011mw}.

Regarding $G_3$ as an independent variable,
the EoM for $G_3$ reads
\begin{align}
  \star G_3=\cF_2 \ ,\nn
\end{align}
reproducing the self-duality condition (\ref{asd:5d}).
We note that the Gauss law constraint for $G_3$
follows from the EoM for $b_1$. 
This guarantees that $G_3$ is regarded as an independent
variable instead of $b_2$.
Integrating out $G_3$ leads to
\begin{align}
  \bar{S}
=\int\bigg[
&-\frac{1}{2}|da|^2-\frac{1}{2}|d\zeta|^2
-\frac{1}{4\pi}\left|\cF_2\right|^2
-\frac{1}{2}\left|i_5\right|^2
-\frac{1}{2}\left|\cF_4\right|^2
-\frac{1}{2}\left|d\phi\right|^2
\nn\\
&
+\frac{M}{4\pi^2}b_1\wedge (da)^2
-\frac{M^2}{16\pi^3}\zeta\,\omega_5
+\frac{K}{2\pi}u_4\wedge d\zeta
+\frac{K}{2\pi}u_3\wedge da
+\frac{N}{16\pi^3}\phi\, d\zeta\wedge(da)^2
\bigg] \ .
\label{5d:action}
\end{align}

The EoM for $b_1$ is
\begin{align}
  d\star\cF_2=\frac{M}{2\pi}(da)^2 \ .
\label{eom:b1}
\end{align}
The EoM for $u_4$ is
\begin{align}
  d\star i_5=-\frac{K}{2\pi}d\zeta \ .
\label{eom:u4}
\end{align}
The EoM for $u_3$ is
\begin{align}
  d\star \cF_4=\frac{K}{2\pi}da \ .
\label{eom:u3}
\end{align}
The EoM for $\phi$ is
\begin{align}
  d\star d\phi=-\frac{N}{16\pi^3}d\zeta\wedge(da)^2 \ .\nn
\end{align}
A bit lengthy computation shows that the EoMs for $\zeta$ and $a$ read
\begin{align}
  d\star d\zeta+\frac{M}{2\pi^2}da\wedge\star\cF_2
-\frac{K}{2\pi}i_5
-\frac{3M^2}{8\pi^2 K}(da)^2\wedge\star\cF_4
-\frac{N}{16\pi^3}d\phi\wedge(da)^2=0 \ ,\nn
\end{align}
and
\begin{align}
  -d\star da
&+\frac{M}{2\pi^2}d\zeta\wedge\star\cF_2
+\frac{M}{2\pi^2}da\wedge\cF_2
-\frac{3M^2}{8\pi^2K}(da)^2\wedge\star i_5
-\frac{3M^2}{4\pi^2 K}d\zeta\wedge da\wedge\star\cF_4
\nn\\
&+\frac{K}{2\pi}\cF_4
+\frac{N}{8\pi^3}d\phi\wedge d\zeta\wedge da
=0 \ ,\nn
\end{align}
respectively. Note that all the EoMs are gauge invariant manifestly.

We discuss the symmetries realized in (\ref{5d:action}) and
their relation to the 6d theory.
(\ref{eom:b1}) gives a conserved current
\begin{align}
  j_{\mathrm{EoM}}^{[1]}=\frac{1}{2\pi}\star\cF_2-\frac{M}{4\pi^2}\omega_3 \ .
\nn
\end{align}
This defines a 1-form symmetry generator.
It is easy to see that the gauge invariance
requires it to generate $\bZ_M^{[1]}$. This is a manifestation
of the 2-form $\bZ_M$ symmetry in 6d after dimensional reduction
to 5d.

The EoM-based conserved current for $\zeta$ and $a$ can be
most easily obtained by dimensionally reducing 
the EoM-based conserved current for $\ha$
\begin{align}
  \hat{j}_{\mathrm{EoM}}^{\prime [1]}=
\hst \hd\ha-\frac{M}{8\pi^2}\,\hd\ha\wedge \hb
-\frac{3M}{4\pi K}\left(
\hd\hb-\frac{M}{2\pi}\,\hat{\omega}_3\right)\wedge
\hst\left(\hd\hu_4+\frac{M^2}{8\pi^2K}\,\hat{\omega}_5\right)
-\frac{K}{2\pi}\,\hu_4-\frac{N}{16\pi^3}\,\hat{\phi}\,(\hd\ha)^2 \ .\nn
\end{align}
to 5d:
\begin{align}
  \hat{j}_{\mathrm{EoM}}^{\prime [1]}=j_{\mathrm{EoM}}^{\prime [0]}
+j_{\mathrm{EoM}}^{\prime [1]}\wedge dX_5 \ .\nn
\end{align}
We find
\begin{align}
  j_{\mathrm{EoM}}^{\prime [0]}&=\star d\zeta-\frac{M}{8\pi^2}da\wedge b_2
-\frac{3M}{4\pi K}\star\cF_2\wedge\star\cF_4
-\frac{K}{2\pi}u_4-\frac{N}{16\pi^3}\phi(da)^2 \ ,
\nn\\
  j_{\mathrm{EoM}}^{\prime [1]}&=\star da-\frac{M}{8\pi^2}\left(da\wedge b_1+d\zeta\wedge b_2\right)
+\frac{3M}{4\pi K}\star\cF_2\wedge\star i_5
-\frac{3M}{4\pi K}\cF_2\wedge\star\cF_4
-\frac{K}{2\pi}u_3-\frac{N}{8\pi^3}\phi d\zeta\wedge da \ .\nn
\end{align}
It is easy to show that $dj_{\mathrm{EoM}}^{\prime [0]}=dj_{\mathrm{EoM}}^{\prime [1]}=0$ by using the EoMs in 5d.
Note that both depend on $b_2$, which is eliminated nonlocally by solving
(\ref{asd:5d}).

Now we examine the gauge invariance condition of
the symmetry generators associated with these currents.
That for $j_{\mathrm{EoM}}^{\prime [0]}$
requires that the corresponding symmetry generator
define $\bZ_q$ with $q$ given in (\ref{qm1N}),
because the CS-like terms in $j_{\mathrm{EoM}}^{\prime [0]}$
take exactly the same form
as in $\hat{j}_{\mathrm{EoM}}^{\prime [1]}$.
{}Furthermore, it is seen 
that the symmetry generator for $j_{\mathrm{EoM}}^{\prime [1]}$ 
is required to 
generate $\bZ_{q}$ as well by examining the difference of
two ways of lifting the CS-like terms in 
$j_{\mathrm{EoM}}^{\prime [1]}$ to a higher-dimensions.


\end{document}